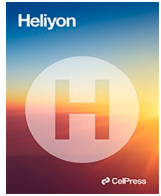

Research article

# Analytical and numerical investigation of heat transfer of porous fin in a local thermal non-equilibrium state

Payam Jalili [a], Salar Ghadiri Alamdari [a], Bahram Jalili [a,*], Amirali Shateri [a], D. D. Ganji [b]

[a] *Department of Mechanical Engineering, Faculty of Engineering, North Tehran Branch, Islamic Azad University, Tehran, Iran*
[b] *Department of Mechanical Engineering, Babol Noshirvani University of Technology, P.O. Box 484, Babol, Iran*

ABSTRACT

This research employs a local thermal non-equilibrium (LTNE) model to analyze the heat transfer phenomenon through a porous fin, considering natural convection and radiation effects. The infiltration velocity within the porous medium is evaluated using the Darcy model, and buoyancy effects are accounted for using the Boussinesq approximation. The Akbari-Ganji method (AGM) is applied to address the governing energy equations. The accuracy of the proposed solution is verified by comparing it with numerical results obtained from the finite difference method (FDM), the finite element method (FEM), and earlier investigations. The results are presented regarding the total average Nusselt number and temperature profiles. These results shed light on the influence of several important parameters, such as the thermal conductivity ratio, dimensionless thickness, convectional heat transfer, and external and internal radiation. The analysis reveals that decreasing Rayleigh and Biot numbers reduces the temperature profiles of the solid phase. Additionally, when the Rayleigh number is low but the assigned Biot number is high, the temperature difference between the solid and fluid phases diminishes. Furthermore, increased thermal conductivity ratio and dimensionless thickness for assigned Biot and Rayleigh numbers lead to higher solid phase temperatures. The Nusselt number exhibits a decreasing trend with a decreasing thermal conductivity ratio but increases with higher Rayleigh and Biot numbers and increased external radiation.

## 1. Introduction

Currently, there is a growing demand for miniaturized devices across various applications. This has sparked significant interest in heat transfer research to achieve optimal thermal performance within the smallest feasible volumes. The pursuit of high heat transfer rates per unit volume has enabled the realization of compact components in diverse technical domains, including solar collectors, automotive engineering, heat exchangers, HVAC systems, refrigeration systems, thermal control in various devices, medical devices, photovoltaic systems, and more. To enhance heat transfer, expanded surfaces and porous media have proven to be effective methods. Porous fins, in particular, offer an advantageous approach by increasing the proportion of heat transfer to the volume. This expanded surface and porous media combination has been applied for various purposes. One of the early studies conducted by Kirwan and Al-Nimr [1] investigated the single-phase porous fin using the finite element method under local thermal equilibrium. The porous

---

medium was modeled using the Brinkman-Forchheimer Darcy model. The study demonstrated a substantial performance improvement when adopting porous fins compared to solid fins, especially at high Darcy numbers with increasing Rayleigh numbers. Another study by Hamdan and Al-Nimr [2] explored the application of porous fins connected to both plates in enhancing forced convection in an isothermal parallel plate duct. The analysis employed the Darcy Brinkman-Forchheimer model and assumed local thermal equilibrium for the porous medium. The results indicated that heat transmission was enhanced with low Darcy number values, highly conductive porous fins, and a higher microscopic inertia coefficient.

Gurla and Bakier [3] conducted an investigation involving three different cases of rectangular porous fins. The study considered the Local Thermal Equilibrium (LTE) hypothesis, employing Darcy's model to characterize the porous medium while accounting for radiation heat transfer. The results revealed that heat transfer due to radiation significantly impacts the case without radiation. In a separate work, Kundu et al. [4] examined the functionality and optimal configuration of straight porous fins with various profiles. The equations of natural convection, based on the Darcy model (Da model), were solved using the Adomian Decomposition Method (ADM). The study demonstrated that porous fins substantially enhance heat transfer compared to solid fins, especially when porosity and flow parameters are relatively low or high. Torabi and Yaghoobi [5] investigated the combined effects of radiative and convective heat transfer on rectangular porous fins. They solved the nonlinear governing equation, resulting from the LTE assumption and Darcy's model for the porous medium, using the Differential Transform Method (DTM). The numerical solution was validated using the fourth-order Runge-Kutta method. The findings were found to be consistent with the results of previous studies.

Hatami et al. [6] studied a rectangular porous fin with temperature-dependent internal heat production and natural convection. The researchers employed the Da model under the LTE assumption. They compared the 4th to 5th-order R-K-Fehlberg method with collocation, differential transform, and least squares solutions to solve the nonlinear governing equation. Their findings revealed the efficacy of the R-K-Fehlberg method for this problem. In related work, Hatami and Ganji [7] investigated the heat transfer due to radiation and natural convection in circular porous fins. They used the LSM (Least Squares Method) and the fourth-order R–K method to solve the governing equations. The study concluded that the exponential profile yielded the highest heat transmission rate. Moradi et al. [8] studied a shifting porous rectangular fin with convection and radiation heat transfer, employing the LTE approach and the Da model theory in their investigation. Additionally, they [9] explored a porous triangular fin with temperature-dependent thermal conductivity. They solved the governing equation using the DTM (Differential Transform Method), comparing the results with those obtained using the fourth-order R–K method. Their findings demonstrated the improved performance of fins when utilizing porous materials. Das [10] conducted numerical research on cylinders with porous fins, investigating heat transfer via convection and radiation. The study utilized the R–K approach to solve the Da model and the governing equation under the LTE assumption. An approach was also employed to define unknown parameters. Rectangular porous fins were examined using the Sumudu transform technique of the Adomian decomposition by Deshamukhya and Meher [11]. The study considered the Da model and the LTE assumption in the governing equation, and their results were found to match those of the HPM (Homotopy Perturbation Method).

Darvishi et al. [12] studied a rectangular porous fin exposed to natural convection and heat transfer by radiation in wet air. The mathematical model utilized the Da method and the LTE assumption, and the spectral collocation method was employed to find the solution. In a different investigation, Motsumi [13] examined radial porous fins with heat transfer involving both radiation and convection, considering temperature-dependent thermal conductivity. The energy equation was solved numerically using the shot technique and the R-K-Fehlberg method, while the nonlinear problem was solved using the LTE assumption and the Da model. Hazarika et al. [14] focused on a porous T-shaped fin with heat and mass transfer due to convection. The study utilized the Da model and the LTE assumption, and the differential transform method was employed to solve the relevant nonlinear governing equations. Ma et al. [15] investigated straight-moving porous fins with various longitudinal profiles, considering radiant heat transmission under the Da hypothesis and LTE assumptions. The mathematical formulation was solved using the Chebyshev polynomial in a spectral approach to address the scenarios of heat generation. Hoseinzadeh et al. [16] studied a convective rectangular porous fin with temperature-dependent heat generation. The researchers used collocation, HPM (Homotopy Perturbation Method), and analytic techniques to solve the nonlinear governing equation, employing the LTE and Da methods.

Kiwan [17] investigated an inclined porous fin with steady heat flow. The governing equation was solved using the Da model and the LTE assumption, providing a sealed response to the base temperature. Hoseinzade et al. [18] studied the thermal behavior of a rectangular porous radiative-convective model with linear heat production. They employed the homotopy analysis approach to resolve the nonlinear problem, considering the Da hypothesis and LTE. Gireesha [19] quantitatively examined a radial porous fin in wet air with natural radiation and convection. The nonlinear central equation, resulting from the LTE and Da hypothesis, was solved using the finite element technique. Shafiei and Talaghat [20] investigated circular porous fins with rectangular, triangular, and convex shapes. The nonlinear energy equation, formulated based on the LTE assumption and the Darcy model, was solved using the Galerkin and finite difference methods. Ndlovu et al. [21] studied moving rectangular porous fins that transfer heat via convection and radiation. The nonlinear equation, formulated under the LTE assumption and the Da model, was solved using the Variational Iteration Method (VIM). Sowmya et al. [22] explored a rectangular porous fin in moist air, considering three different temperature-dependent thermal conductivity values. Based on Darcy's formula and LTE, the nonlinear energy equation was resolved using the 4th and 5th-order R-K-Fehlberg method. Buonomo et al. [23] investigated heat transfer in rectangular porous fins using a local thermal non-equilibrium model. The nonlinear governing equation, formulated under the Darcy and local thermal non-equilibrium models, was solved using the Adomian decomposition method. Manohar et al. [24] studied the thermal behavior of ethylene, water, glycol, and motor oil transporting SWCNT-MWCNT nanoparticles in a magnetic field and heat production. The heat equation was developed using the Darcy law concept. The non-dimensional governing equation was solved using Maple software and the 4th and 5th orders of the R–K Fehlberg approach. The study showed that convection and radiation factors decreased thermal performance, while the magnetic constraint and internal heat generation parameters enhanced it. Ethylene glycol outperformed water and motor oil-based liquids regarding heat distribution when SWCNT-MWCNT was present.

Sowmya et al. [25] investigated the thermal behavior of a porous horizontal fin in a moisture environment with heat transfer effects. They considered the fin's thickness varying with length, which led to examining various fin profiles, including triangular, rectangular, and convex fins. The Da model was used to account for the fin's porousness. The numerical solution was obtained using the finite difference method and Maple software to solve the resulting nonlinear partial differential equation. The study revealed that rectangular fins are more efficient in heat transfer, while triangular fins exhibit a faster rate of temperature decline. In another study, Sowmya and Gireesha [26] considered radiation, natural convection, and thermally dependent internal heat generation in the analysis of a porous fin with a longitudinal geometry having diverse profiles such as rectangular, triangular, and convex shapes in a thoroughly wet condition. Aluminum and copper metals were used to create the fin. The Differential Transform Method (DTM) was employed to solve the resulting ordinary differential equation. The findings showed that the aluminum fin dissipates heat more effectively than the copper fin. Hosseinzadeh et al. [27] studied the thermal performance of a moving porous fin immersed in a hybrid nanofluid. They considered concave parabolic, convex, and trapezoidal cross-sections for the fin. After deriving the governing equations and transforming the partial differential equations to ordinary differential equations using the Similarity solution, they solved the equations using Akbari-Ganji's approach. The study demonstrated that Akbari-Ganji's approach is reliable for resolving heat transfer equations involving moving porous fins.

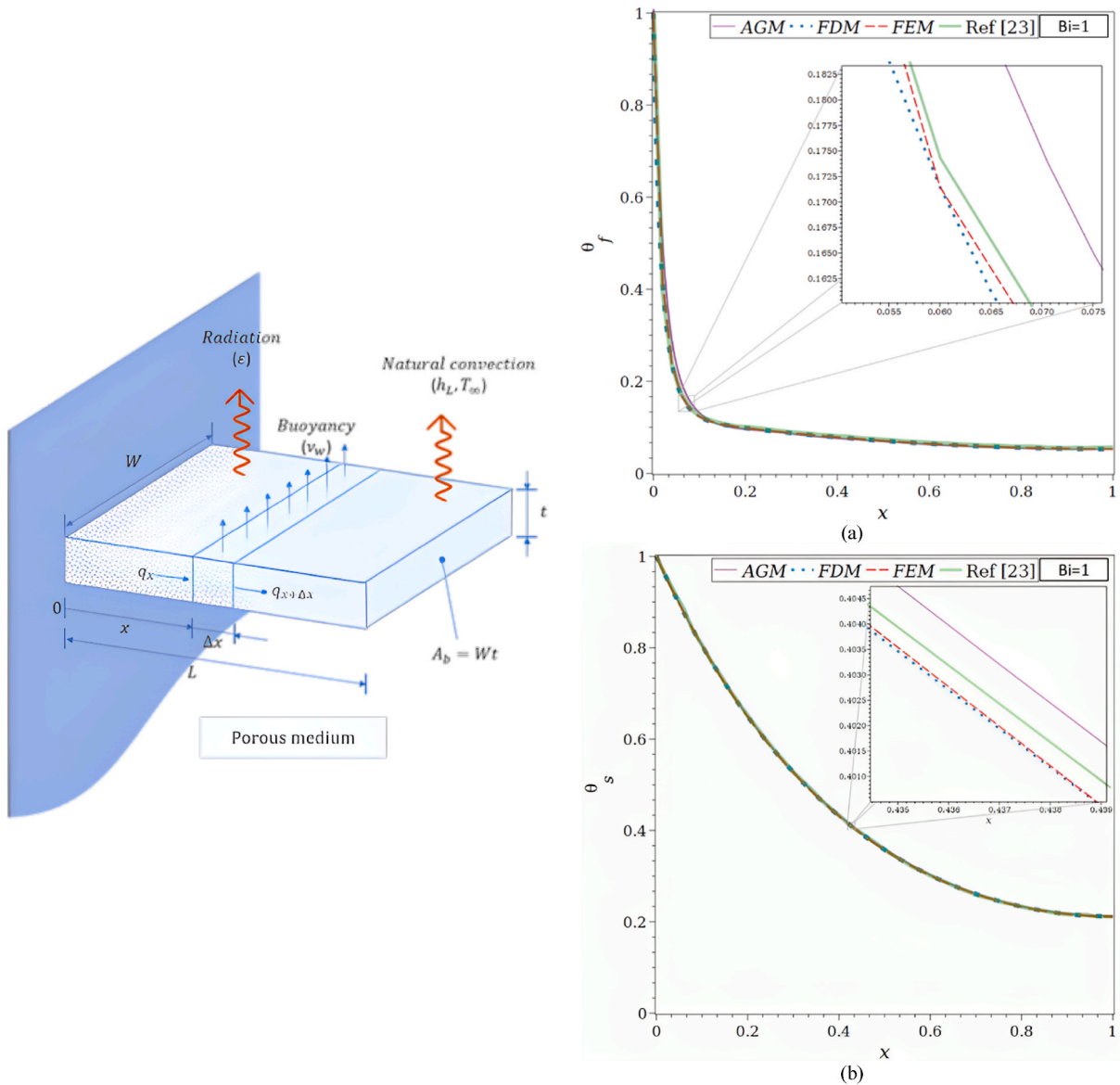

**Fig. 1.** a. The porous fin's geometry.
Fig. 1b. Comparison of the solutions by AGM, FDM, FEM, and ref [23] for Bi = 1, Ra = 50, $\tau = 0.1$, $R_d = 0.1$, $\lambda = 0.01$, $R_1 = 0.3$, $\varphi = 0.92$, $\kappa = 1000$: (a) $\theta_f$, (b) $\theta_s$.
Fig. 1c. Comparison of the solutions by AGM, FDM, FEM, and ref [23] for Bi = 100, Ra = 50, $\tau = 0.1$, $R_d = 0.1$, $\lambda = 0.01$, $R_1 = 0.3$, $\varphi = 0.92$, $\kappa = 1000$: (a) $\theta_f$, (b) $\theta_s$.
Fig. 1d. Function of Ra for the total Nusselt number by AGM, FDM, FEM, and ref [23] for Ra = 50, $\tau = 0.1$, $R_d = 0.1$, $\lambda = 0.01$, $R_1 = 0.3$, $\varphi = 0.92$, $\kappa = 1000$: (a) Bi = 1, (b) Bi = 100.

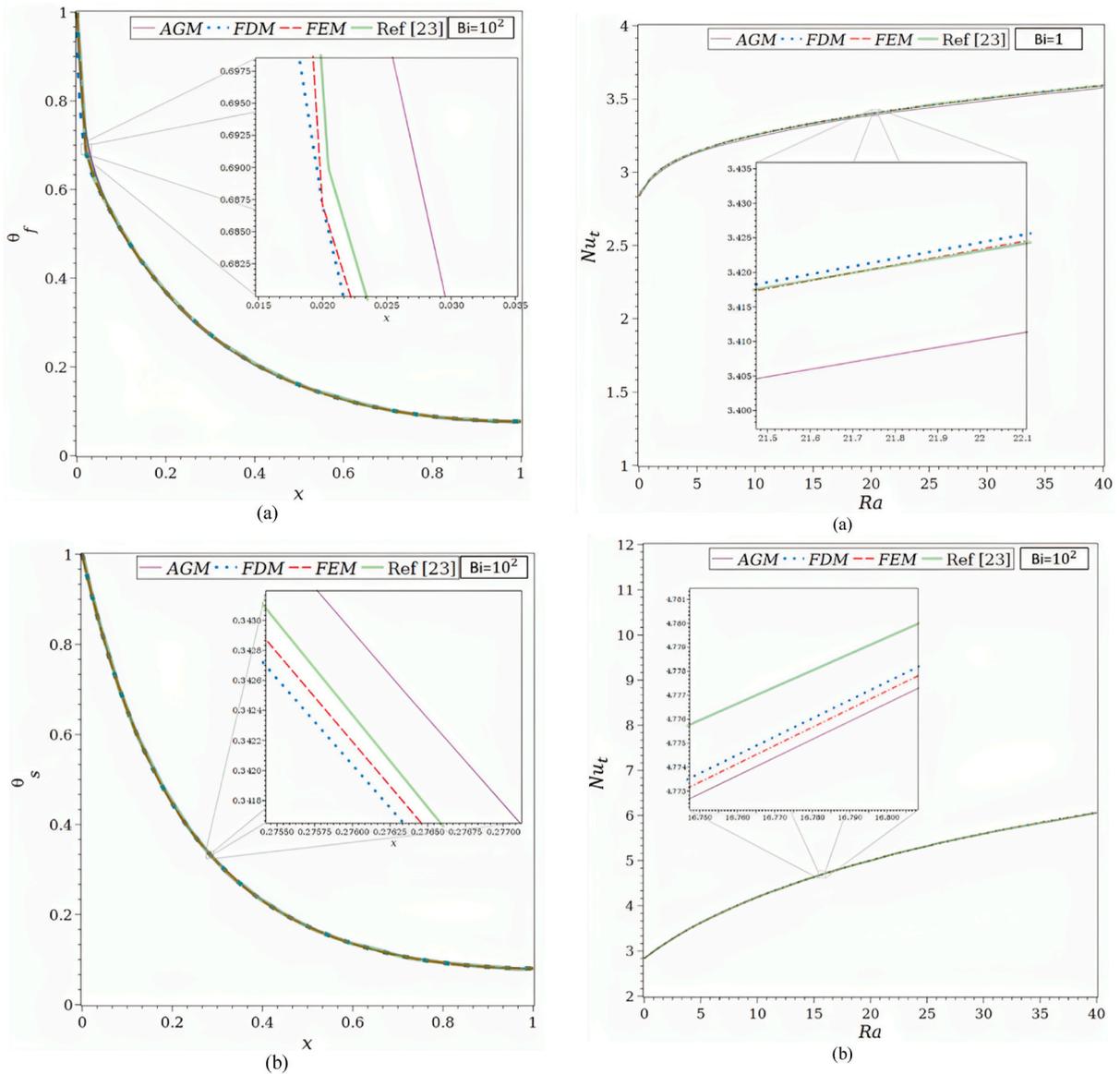

**Fig. 1.** (*continued*).

In the study by Sowmya and Gireesha [28], they investigated a porous fin with a radial profile while considering internal heat generation that is temperature-dependent and subject to radiation. Two scenarios of thermal conductivity were examined: one where it varies linearly with temperature and another where it varies exponentially. The fin's porousness was accounted for using Darcy's law. The resulting equation was a second-order nonlinear ordinary differential equation (ODE), solved using the R-K-Fehlberg 4th and 5th-order method. The study's findings showed that as the Peclet number, a dimensionless parameter representing the ratio of conduction to convection heat transfer, increases, the fin moves more quickly, leading to higher dimensionless temperature. Additionally, the study revealed that the tangential stress exhibits more tensile stress towards the tip radius and more significant compression near the fin base region.

Kumar et al. [29] used the DTM-Pade approximant algorithm to solve a nonlinear ordinary differential equation resulting from non-dimensionalizing the heat equation using non-dimensional terms. The main results of their study indicated that thermal distribution decreased with increasing heat transfer parameters, radiation-conduction parameters, and fin taper ratio values. Ullah et al.

[30] researched a moving convection-radiative porous fin subjected to heat production. They analyzed the model both numerically and analytically to investigate strategies for increasing efficiency. The study examined the effects of various parameters on the temperature distribution and performance of the fins. In another study, Wang et al. [31] investigated an inclined porous longitudinal fin subjected to convective and radiative phenomena. The Pade approximant and differential transform method (DTM) were used to solve the nonlinear heat transfer system. The findings showed that an increase in the conduction-convection variable, porosity variable, and radiation number decreased the thermal field. Additionally, the thermal behavior of the fin decreased with a higher degree of inclination. Weera et al. [32] studied the convective-radiative thermal behavior of a longitudinal dovetail porous fin with internal heat generation. The heat transfer issue was governed by a second-order ODE, which was numerically solved using the spectral collocation technique with a local linearization strategy. Further applications of Akbari Ganji's method can be found in Refs. [33–40]. Also, Nandeppanavar et al. [41] investigated the unsteady MHD stream of Casson fluid over an elongating surface. Their study considered the effects of thermal radiation and viscous dissipation on the flow behavior. The authors presented numerical results for the velocity and temperature profiles, highlighting the impact of these factors on the flow characteristics. In a related study, Nandeppanavar et al. [42] numerically examined the effect of the Richardson number on the stagnation point flow of double-diffusive mixed convective slip flow of magnetohydrodynamic Casson fluid. Their findings revealed the significant influence of the Richardson number on the flow and heat transfer characteristics, providing valuable insights into the behavior of such flows. Mathematical modeling of engineering problems involving non-Newtonian fluids has also been a topic of interest. Kemparaju et al. [43] presented a mathematical modeling approach for engineering problems involving Casson fluid. Their study focused on developing mathematical models and numerical techniques to analyze and solve engineering problems related to non-Newtonian fluid flows. This work contributes to understanding the application of Casson fluid in engineering disciplines. In an earlier study, Nandeppanavar et al. [44] investigated the slip flow and heat transfer due to a stretching sheet in the presence of Casson fluid. The authors derived an analytical solution for the velocity and temperature profiles, providing a valuable tool for understanding the behavior of these flows.

This research aims to fill the gap in the existing literature by applying the Akbari-Ganji method (AGM) to analyze heat transport in a porous fin under a local thermal non-equilibrium state. To our knowledge, this method has not been utilized for such an investigation before. The study seeks to validate its findings using various sources and numerical methods.

## 2. Governing equations

As depicted in Fig. 1a, the physical issue being studied involves a vertical wall and a convection-driven rectangular porous fin and radiation heat transmission. The porous fin's cross-section area, $A_b = W \times t$ is taken to be constant along the longitudinal axis x. Its width and thickness are W and t, respectively. The fin is L in length. Since the fin is porous, fluid can travel through it as a result of the buoyancy force produced by the base surface's constant temperature, $T_b$, which is where the porous fin is located.

The analysis is done with the following presumptions in mind:

- The porous material is homogeneous, isotropic, and filled with a single-phase fluid;
- Except for the density in the buoyancy term, all of the solid and the fluid matrix's thermophysical properties are temperature-independent (Boussinesq approximation);
- Heat transfer happens along the x-axis, and the temperature within a porous fin is just a function of the longitudinal coordinate (x) along the fin's length L;
- The Da method represents the momentum between the fluid and the porous fin. The buoyancy effect produces fluid motion orthogonal to the fin length L.
- Radiative interactions between the environment and the porous fin's outer surface are considered, and a constant emissivity is assumed.
- Heat sources within the fin and thermal contact resistances at their bases have all been adjusted to zero.
- The local thermal non-equilibrium theory is considered to simulate the thermal relation between porous fin and fluid. This supposition states that local variations in solid and fluid temperatures exist.
- The volumetric and surface porosities are equal.

These theories lead to the following equations of energy for solid and fluid phases in a steady state; Fluid phase:

$$\dot{m}c_{pf}\left(T_f - T_\infty\right) = h_{sf}A_{sf}\left(T_s - T_f\right) + A_b\dot{q}_f(x) - A_b\dot{q}_f(x + \Delta x) \tag{1}$$

Solid phase:

$$A_b\dot{q}_s(x) - A_b\dot{q}_s(x + \Delta x) = h_{sf}A_{sf}\left(T_s - T_f\right) + h_L P\Delta x(1-\varphi)(T_s - T_\infty) + P\Delta x\sigma\varepsilon\left(T_s^4 - T_\infty^4\right) \tag{2}$$

where, respectively, $T_f$ represents the fluid phase and $T_s$ represents the solid phase, $T_\infty$ is ambient temperatures; $\varphi$ defines as the porosity; $\varepsilon$ is the emissivity coefficient; $\sigma$ defines as the Stefan-Boltzmann constant; $h_{sf}$ and $A_{sf}$ define as the convective heat transfer coefficient and the internal porous fin's interfacial area, respectively, whereas $h_L$, is the exterior convective heat transfer coefficient.

The formula for the buoyant force's impact on the fluid's mass flow rate is [1]:

$$\dot{m} = \rho_f(W \bullet \Delta x)v_w \tag{3}$$

where $\nu_w$, is the fluid's velocity through the porous medium due to buoyancy. The Darcy model can be used to assess the infiltration velocity, $\nu_w$, as shown in:

$$v_w = \frac{gK\beta}{\nu}\left(T_f - T_\infty\right) \tag{4}$$

The following results from substituting Equations (3) and (4) for Equations (1) and (2) and fractionalizing by the term $\Delta x \bullet A_b$:

$$\frac{\acute{q_f}(x) - \acute{q_f}(x + \Delta x)}{\Delta x} = \frac{h_{sf}A_{sf}}{A_b\Delta x}\left(T_f - T_s\right) + \frac{\rho_f c_{pf} gK\beta}{\nu t}\left(T_f - T_\infty\right)^2 \tag{5}$$

$$\frac{\acute{q_s}(x) - \acute{q_s}(x + \Delta x)}{\Delta x} = \frac{h_{sf}A_{sf}}{A_b\Delta x}\left(T_s - T_f\right) + \frac{h_L P(1-\varphi)}{A_b}\left(T_s - T_\infty\right) + \frac{P\sigma\varepsilon}{A_b}\left(T_s^4 - T_\infty^4\right) \tag{6}$$

For, $\Delta x \to 0$, Eqs. (5) and (6) become:

$$-\frac{d\acute{q_f}(x)}{dx} = h_{sf}a_{sf}\left(T_f - T_s\right) + \frac{\rho_f c_{pf} gK\beta}{\nu t}\left(T_f - T_\infty\right)^2 \tag{7}$$

$$-\frac{d\acute{q_s}(x)}{dx} = h_{sf}a_{sf}\left(T_s - T_f\right) + \frac{h_L P(1-\varphi)}{A_b}\left(T_s - T_\infty\right) + \frac{P\sigma\varepsilon}{A_b}\left(T_s^4 - T_\infty^4\right) \tag{8}$$

where $h_{sf}$, is the convective heat transfer coefficient, and $a_{sf} = A_{sf}/(A_b\Delta x)$ is the interfacial area per unit of volume in the porous fin. $a_{sf}$ and $h_{sf}$, related to the porous media, and available literature evaluates both [45].

The Rosseland diffusion model to calculate the effects of internal radiation (Equation (9)):

$$\acute{q_f}(x) = -\varphi k_f\frac{dT_f}{dx}, \acute{q_s}(x) = -(1-\varphi)\,k_s\frac{dT_s}{dx} - \frac{4\sigma}{3\beta_R}\frac{dT_s^4}{dx} \tag{9}$$

Rosseland's mean extinction coefficient is represented by $\beta_R$.

Eqs. (7) and (8) can be stated as follows by linearized formulation of $T_s^4$ around ambient temperature $T_\infty$:

$$\varphi k_f\frac{dT_f}{dx} = h_{sf}a_{sf}\left(T_f - T_s\right) + \frac{\rho_f c_{pf} gK\beta}{\nu t}\left(T_f - T_\infty\right)^2 \tag{10}$$

$$\left[(1-\varphi)\,k_s + \frac{16\sigma T_\infty^3}{3\beta_R}\right]\frac{d^2 T_s}{dx^2} = h_{sf}a_{sf}\left(T_s - T_f\right) + \left[\frac{h_L P(1-\varphi)}{A_b} + \frac{4P\sigma\varepsilon T_\infty^3}{A_b}\right]\left(T_s - T_\infty\right) \tag{11}$$

Equations 10 and 11 represent a collection of 2nd-order nonlinear ODEs. The B.C.s for Equations (10) and (11) are as follows:

$$T_f(0) = T_s(0) = T_b \text{ at base fin } \frac{dT_f}{dx}(L) = \frac{dT_s}{dx}(L) = 0 \text{ at tip fin} \tag{12}$$

By employing dimensionless variables as (Equation (13)):

$$X = \frac{x}{L}; \theta = \frac{T - T_\infty}{T_b - T_\infty}; \tau = \frac{t}{L}; \frac{PL}{A_b} \simeq \frac{2}{\tau}; R_d = \frac{4\sigma T_\infty^3}{3\beta_R(1-\varphi)k_s}; Ra = \frac{gK\beta(T_b - T_\infty)t}{\varphi\nu_f a_f}; \lambda = \frac{h_L L}{k_s}; R_1 = \frac{4\sigma\varepsilon T_\infty^3 L}{(1-\varphi)k_s}; Bi = \frac{h_{sf}a_{sf}L^2}{(1-\varphi)k_s}; \chi = \frac{(1-\varphi)}{\varphi}\kappa; \kappa = \frac{k_s}{k_f} \tag{13}$$

The boundary conditions (equation (12)) and the energy equations 10 and 11 can be expressed as:

$$\frac{d^2\theta_f}{dX} = \chi Bi\left(\theta_f - \theta_s\right) + \frac{Ra}{\tau^2}\theta_f^2 \tag{14}$$

$$[1 + 4R_d]\frac{d^2\theta_s}{dX} = Bi\left(\theta_s - \theta_f\right) + \frac{2}{\tau}(\lambda + R_1)\theta_s \tag{15}$$

$$\theta_f(0) = \theta_s(0) = 1; \frac{d\theta_f}{dx}(1) = \frac{d\theta_s}{dx}(1) = 0 \tag{16}$$

And:

The following definitions apply to both the fluid and solid matrix average Nusselt numbers (Equation (17) and (18)):

$$Nu_f = \frac{\acute{q_{f,b}}t}{k_{eq}(T_b - T_\infty)} = -\frac{\varphi\tau}{\varphi + (1-\varphi)\kappa}\acute{\theta_f}(0) \tag{17}$$

$$Nu_s = \frac{\acute{q_{s,b}}t}{k_{eq}(T_b - T_\infty)} = -\frac{(1-\varphi)\kappa\tau}{\varphi + (1-\varphi)\kappa}(1 + 4R_d)\acute{\theta_s}(0) \tag{18}$$

Moreover, the porous fin's total Nusselt number is (Equation (19)):

$$Nu_t = \frac{(\dot{q}_{f,b} + \dot{q}_{s,b})t}{k_{eq}(T_b - T_\infty)} = Nu_f + Nu_s \tag{19}$$

Where the fluid and solid heat fluxes are $\dot{q}_{f,b}$ and $\dot{q}_{s,b}$, respectively, at the base of the fin. According to Ref. [45], the porous medium affects the value of $k_{eq}$. As stated in Ref. [1], the effective thermal conductivity of porous fin is adopted in the current inquiry as $k_{eq} = \varphi k_f + (1 - \varphi)k_s$.

## 3. Methodology

### 3.1. Akbari-Ganji's method

The general approach to a differential equation depends on the boundary conditions and is as follows:

$$p_k : f\left(u, u^{'}, u^{''}, \ldots, u^{(m)}\right) = 0; u = u(x) \tag{20}$$

The parameter u, which is a function of x, and the nonlinear differential equation of p, which is a function of u, and their derivatives are as follows:
Boundary conditions:

$$\begin{cases} u(x) = u_0, u^{'}(x) = u_1, \ldots, u^{(m-1)}(x) = u_{m-1} & \text{at } x = 0 \\ u(x) = u_{L_0}, u^{'}(x) = u_{L_1}, \ldots, u^{(m-1)}(x) = u_{L_{m-1}} & \text{at } x = L \end{cases} \tag{21}$$

We presume that the solution of the first differential equation is taken into consideration to solve the first differential equation for the B.C.s in $x = L$ in Eq. (21) as follows:

$$u(x) = \sum_{i=0}^{n} a_i x^i = a_0 + a_1 x^1 + a_2 x^2 + \ldots + a_n x^n \tag{22}$$

The additional series statements in Eq. (22) lead to a more accurate solution for Eq. (20). There are (n+1) unknown coefficients that require the specification of (n+1) equations to achieve the solution of the differential equation (20) corresponding to the series from degree (n).
For the solution of the differential equation (22), the boundary conditions are applied as follows:
When x = 0:

$$\begin{cases} u(0) = a_0 = u_0 \\ u^{'}(0) = a_0 = u_0 \\ u^{''}(0) = a_2 = u_2 \\ \quad \vdots \quad \vdots \quad \vdots \end{cases} \tag{23}$$

And when x = L:

$$\begin{cases} u(L) = a_0 + a_1 L + a_2 L^2 + \ldots + a_n L^n = u_{L_0} \\ u^{'}(L) = a_1 + 2a_2 L + 2a_3 L^2 + \ldots + na_n L^{n-1} = u_{L_1} \\ u^{''}(L) = 2a_2 + 6a_3 L + 12a_4 L^2 + \ldots + n(n-1)a_n L^{n-2} = u_{L_{m-1}} \\ \quad \vdots \quad \vdots \quad \vdots \quad \vdots \quad \vdots \quad \vdots \quad \vdots \end{cases} \tag{24}$$

Following the substitution of Eq. (24) into Eq. (20), the boundary conditions are applied to differential Eq. (20) in the manner shown below (Equation (25)):

$$p_0 : f\left(u(0), u^{'}(0), u^{''}(0), \ldots, u^{(m)}(0)\right) \quad p_1 : f\left(u(L), u^{'}(L), u^{''}(L), \ldots, u^{(m)}(L)\right) \tag{25}$$

We must select n; (n < m) phrases from Eq. (22) and come across many additional unknowns that are, in reality, the same co-efficients of Eq. (22). To resolve this problem, we must first apply the B.C.s to the differential equations mentioned above before calculating m using the additional unknowns from Eq. (20).

$$p_k^{'} : f\left(u^{'}, u^{''}, u^{'''}, \ldots, u^{(m+1)}\right) \quad p_k^{''} : f\left(u^{''}, u^{'''}, u^{(IV)}, \ldots, u^{(m+2)}\right) \vdots \vdots \vdots \vdots \vdots \tag{26}$$

The B in equation (26) is applied to the derivatives of the $P_k$ , as follows (Equation (27) and (28)):

$$p_k' : \begin{cases} f\big(u'(0), u''(0), u'''(0), \ldots, u^{(m+1)}(0)\big) \\ f\big(u'(L), u''(L), u'''(L), \ldots, u^{(m+1)}(L)\big) \end{cases} \tag{27}$$

$$p_k'' : \begin{cases} f\big(u'(0), u''(0), u^{(IV)}(0), \ldots, u^{(m+2)}(0)\big) \\ f\big(u'(L), u''(L), u^{(IV)}(L), \ldots, u^{(m+2)}(L)\big) \end{cases} \tag{28}$$

To calculate the $(n+1)$ unknown coefficients of Eq. (22), $(n+1)$ equations built from equations (23)–(28). Finding the coefficients in equation (22) will lead to the answer to the nonlinear differential equation (20).

## 3.2. Application of AGM

Assuming the following exponential and polynomial with constant coefficients as the solution to equations (14) and (15) will allow us to apply the AGM procedure outlined before:

$$\theta_f = \sum_{i=0}^{22} a_i e^{-0.86ix} \tag{29}$$

$$\theta_s = \sum_{i=0}^{6} b_i x^i \tag{30}$$

According to the AGM description, we assume 23 sentences for $\theta_f$ and seven sentences for $\theta_s$. So, there are 30 unknowns based on equations (29) and (30). To obtain 30 unknowns, we need 30 equations. By applying the boundary conditions in a series solution of $\theta_f$, two equations have been made. The rest of equation (21 equations) should be made by applying the AGM. Simultiniously, to determine the unknowns of $\theta_s$, we need seven equations. Boundary conditions give us two equations; another five equations have been created according to the AGM. The system of ODEs (14 and 15) with the parameters Ra = 50, $\tau = 0.1$, $R_d = 0.1$, $\lambda = 0.01$, $R_1 = 0.3$, $\varphi = 0.92$, $\kappa = 1000$ can be solved by computing the system of 30 unknowns and 30 equations, and the following functions are the solutions (Equation (31)–(34)):

For Bi = 1:

$$\begin{aligned} \theta_f &= 842.8191124 - 11576.58148\, e^{-0.86x} + 62318.74159\, e^{-1.72x} - 153706.4734\, e^{-2.58x} + 152840.8065\, e^{-3.44x} - 362395.9694\, e^{-4.30x} \\ &\quad + 3.151398792 \times 10^6 e^{-5.16x} - 1.167675907 \times 10^7 e^{-6.02x} + 2.254342754 \times 10^7 e^{-6.88x} - 2.590381477 \times 10^7 e^{-7.74x} + 2.274808032 \\ &\quad \times 10^7 e^{-8.60x} - 2.451164619 \times 10^7 e^{-9.46x} + 2.083063993 \times 10^7 e^{-10.32x} + 9.640081266 \times 10^6 e^{-11.18x} - 4.207652209 \times 10^7 e^{-12.04x} \\ &\quad + 3.151594098 \times 10^7 e^{-12.90x} + 5.527905467 \times 10^6 e^{-13.76x} - 2.117488043 \times 10^7 e^{-14.62x} + 1.498938114 \times 10^7 e^{-15.48x} - 1.060974569 \\ &\quad \times 10^7 e^{-16.34x} + 8.895356239 \times 10^6 e^{-17.20x} - 4.448627781 \times 10^6 e^{-18.06x} + 871462.0328\, e^{-18.92x} \end{aligned} \tag{31}$$

$$\theta_s = 1.0 - 2.171515552x + 0.03829673215x^2 - 0.2297805627x^3 + 0.7961210685x^4 - 1.652613261x^5 + 2.431996741x^6 \tag{32}$$

For Bi = 100:

$$\begin{aligned} \theta_f &= 11170.44798 - 170714.4898\, e^{-0.86x} + 1.080573768 \times 10^6 e^{-1.72x} - 3.495237252 \times 10^6 e^{-2.58x} + 5.298534484 \times 10^6 e^{-3.44x} \\ &\quad - 453759.6765\, e^{-4.30x} - 6.016909675 \times 10^6 e^{-5.16x} - 1.512030518 \times 10^7 e^{-6.02x} + 7.009479783 \times 10^7 e^{-6.88x} - 8.950141697 \\ &\quad \times 10^7 e^{-7.74x} + 3.229204365 \times 10^7 e^{-8.60x} + 5.457277055 \times 10^6 e^{-9.46x} + 1.774850235 \times 10^7 e^{-10.32x} + 3.225231255 \times 10^7 e^{-11.18x} \\ &\quad - 1.384597827 \times 10^8 e^{-12.04x} + 8.624010958 \times 10^7 e^{-12.90x} + 6.573252777 \times 10^7 e^{-13.76x} - 5.993397159 \times 10^7 e^{-14.62x} - 5.159081738 \\ &\quad \times 10^7 e^{-15.48x} + 6.460591230 \times 10^7 e^{-16.34x} - 7.452683599 \times 10^6 e^{-17.20x} - 1.282336469 \times 10^7 e^{-18.06x} + 4.205202458 \times 10^6 e^{-18.92x} \end{aligned} \tag{33}$$

$$\begin{aligned} \theta_s &= 1.002504934 - 4.42352990435958x + 11.3857089312886x^2 - 19.4852112351277x^3 + 21.1572104583124x^4 \\ &\quad - 12.8368937323794x^5 + 3.28005807473363x^6 \end{aligned} \tag{34}$$

The solution functions for other sets of different parameters can be obtained using the same technique.

## 3.3. Numerical methods

The two numerical techniques used to verify the findings reached by AGM are the finite difference method (FDM) and the finite element method (FEM). Equations (14) and (15) have been solved using both methods, with equation (16) serving as the boundary condition.

**Table 1**

Analogy of the $\theta_f$ and $\theta_s$ by AGM, FDM, FEM, and ref [23] for Bi = 1, Ra = 50, $\tau$ = 0.1, $R_d$ = 0.1, $\lambda$ = 0.01, $R_1$ = 0.3, $\varphi$ = 0.92, $\kappa$ = 1000.

| X | $\theta_f$ | | | | $\theta_s$ | | | |
|---|---|---|---|---|---|---|---|---|
| | AGM | FDM | FEM | Ref [23] | AGM | FDM | FEM | Ref [23] |
| 0.0 | 1.0000000000 | 1.0000000000 | 1.0000000000 | 1.0000000000 | 1.0000000000 | 1.0000000000 | 1.0000000000 | 1.0000000000 |
| 0.1 | 0.1317803060 | 0.1233851696 | 0.1233851824 | 0.1248097412 | 0.8055931515 | 0.8044731357 | 0.8044731359 | 0.8052805281 |
| 0.2 | 0.0961488253 | 0.0985431013 | 0.0985431017 | 0.1004566210 | 0.6509585681 | 0.6494236985 | 0.6494236988 | 0.6509900990 |
| 0.3 | 0.0860212731 | 0.0876181305 | 0.0876181305 | 0.0890410959 | 0.5287226152 | 0.5272043522 | 0.5272043525 | 0.5280528053 |
| 0.4 | 0.0762416715 | 0.0785323195 | 0.0785323196 | 0.0806697108 | 0.4329306189 | 0.4315852403 | 0.4315852406 | 0.4323432343 |
| 0.5 | 0.0695714346 | 0.0708223640 | 0.0708223639 | 0.0730593607 | 0.3588400623 | 0.3576930599 | 0.3576930601 | 0.3580858086 |
| 0.6 | 0.0628852075 | 0.0644529728 | 0.0644529727 | 0.0662100457 | 0.3027413574 | 0.3017666522 | 0.3017666523 | 0.3011551155 |
| 0.7 | 0.0580113911 | 0.0594350943 | 0.0594350944 | 0.0616438356 | 0.2618061898 | 0.2609631384 | 0.2609631385 | 0.2590759076 |
| 0.8 | 0.0546865156 | 0.0558027650 | 0.0558027650 | 0.0578386606 | 0.2339634364 | 0.2332110183 | 0.2332110185 | 0.2334983498 |
| 0.9 | 0.0524845998 | 0.0535991733 | 0.0535991732 | 0.0547945205 | 0.2178026583 | 0.2171029430 | 0.2171029431 | 0.2169966997 |
| 1.0 | 0.0527201404 | 0.0528600331 | 0.0528600325 | 0.0563165906 | 0.2125051660 | 0.2118227266 | 0.2118227266 | 0.2112211221 |

**Table 2**

Analogy of the $\theta_f$ and $\theta_s$ by AGM, FDM, FEM, and ref [23] for Bi = 100, Ra = 50, $\tau = 0.1$, $R_d = 0.1$, $\lambda = 0.01$, $R_1 = 0.3$, $\varphi = 0.92$, $\kappa = 1000$.

| X | $\theta_f$ | | | | $\theta_s$ | | | |
|---|---|---|---|---|---|---|---|---|
| | AGM | FDM | FEM | Ref [23] | AGM | FDM | FEM | Ref [23] |
| 0.0 | 0.9977353330 | 1.0000000000 | 1.0000000000 | 1.0000000000 | 1.0025049341 | 1.0000000000 | 1.0000000000 | 1.0000000000 |
| 0.1 | 0.5050673900 | 0.5072102227 | 0.5072102243 | 0.5083333333 | 0.6565144539 | 0.6543906114 | 0.6543906144 | 0.6543683589 |
| 0.2 | 0.3665035900 | 0.3686506850 | 0.3686506863 | 0.3691666667 | 0.4472992750 | 0.4463076136 | 0.4463076156 | 0.4486639697 |
| 0.3 | 0.2711839140 | 0.2731554230 | 0.2731554238 | 0.2725000000 | 0.3166299785 | 0.3157326028 | 0.3157326040 | 0.3160433595 |
| 0.4 | 0.2046486940 | 0.2064149968 | 0.2064149968 | 0.2066666667 | 0.2313627961 | 0.2306908384 | 0.2306908393 | 0.2306598259 |
| 0.5 | 0.1598496720 | 0.1593563632 | 0.1593563684 | 0.1608333333 | 0.1739394422 | 0.1738001809 | 0.1738001815 | 0.1737967914 |
| 0.6 | 0.1263532640 | 0.1262062697 | 0.1262062703 | 0.1275000000 | 0.1352485882 | 0.1352484878 | 0.1352484882 | 0.1347605158 |
| 0.7 | 0.1028346823 | 0.1032874880 | 0.1032874883 | 0.1033333333 | 0.1098489776 | 0.1093316302 | 0.1093316305 | 0.1090917772 |
| 0.8 | 0.0877568794 | 0.0882803695 | 0.0882803697 | 0.0883333333 | 0.0935541837 | 0.0926874045 | 0.0926874048 | 0.0932269602 |
| 0.9 | 0.0784937227 | 0.0797770231 | 0.0797770233 | 0.0800000000 | 0.0833790090 | 0.0833709179 | 0.0833709181 | 0.0844922680 |
| 1.0 | 0.0754651688 | 0.0770218934 | 0.0770218936 | 0.0775000000 | 0.0798475265 | 0.0803700852 | 0.0803700854 | 0.0802139037 |

**Table 3**

Analogy of $Nu_t$ as a function of Ra by AGM, FDM, FEM, and ref [23] for Ra = 50, $\tau = 0.1$, $R_d = 0.1$, $\lambda = 0.01$, $R_1 = 0.3$, $\varphi = 0.92$, $\kappa = 1000$.

| Ra | $Nu_t$ | | | | | | | |
|---|---|---|---|---|---|---|---|---|
| | Bi = 1 | | | | Bi = $10^2$ | | | |
| | AGM | FDM | FEM | Ref [23] | AGM | FDM | FEM | Ref [23] |
| 0 | 2.8581612941 | 2.8370370865 | 2.8370368620 | 2.8376511226 | 2.8351767143 | 2.8374895652 | 2.8374893410 | 2.8377926421 |
| 5 | 3.1326426944 | 3.1413322314 | 3.1412722510 | 3.1381692573 | 3.6268798353 | 3.6268932306 | 3.6268588740 | 3.6304347826 |
| 10 | 3.2381540780 | 3.2550266016 | 3.2548027990 | 3.2538860104 | 4.1914329366 | 4.1919134961 | 4.1917789960 | 4.1973244147 |
| 15 | 3.3210612679 | 3.3357376745 | 3.3352620910 | 3.3350604491 | 4.6364001456 | 4.6374943161 | 4.6371962010 | 4.6387959866 |
| 20 | 3.3882058668 | 3.4010065498 | 3.4002016490 | 3.4024179620 | 5.0063042488 | 5.0068011843 | 5.0062789080 | 4.9949832776 |
| 25 | 3.4402927316 | 3.4570074342 | 3.4558033480 | 3.4594127807 | 5.3247187671 | 5.3226036921 | 5.3217996470 | 5.3210702341 |
| 30 | 3.4890830683 | 3.5067080770 | 3.5050409750 | 3.5060449050 | 5.6002427914 | 5.5985977955 | 5.5974571930 | 5.5969899666 |
| 35 | 3.5345802131 | 3.5517890219 | 3.5496000290 | 3.5492227979 | 5.8340702866 | 5.8437367376 | 5.8422074400 | 5.8377926421 |
| 40 | 3.5783951688 | 3.5933051946 | 3.5905396030 | 3.5906735751 | 6.0668655647 | 6.0642240155 | 6.0622563610 | 6.0585284281 |

### 3.3.1. Finite difference method

The numerical solutions of differential equations based on finite differences give us the values at discrete grid points. The concept of FDM is to replace the derivatives found in the governing differential equations with algebraic difference approaches. In order to find the values of the dependent variables at the discrete grid points, this creates an algebraic equation system that can be solved using any standard analytical or numerical methods.

The rate of change of the function to the variable $x$ is expressed using the present value at $x = x_i$ and the forward step at $x_{i+1} = x_i + \Delta x_i$. Mathematically, the derivative of the function $(x)$, corresponds to this (Equation (35)):

$$\Delta \theta_i = \frac{\mathrm{d}\theta(x)}{\mathrm{d}x}\bigg|_{x=x_i} \approx \frac{\theta_{i+1} - \theta_i}{x_{i+1} - x_i} = \frac{\theta_{i+1} - \theta_i}{\Delta x} = \frac{\theta_{i+1} - \theta_i}{h}, \Delta \theta_{i+1} = \frac{\theta_{i+2} - \theta_{i+1}}{h}, \Delta \theta_{i+2} = \frac{\theta_{i+3} - \theta_{i+2}}{h}, etc. \tag{35}$$

Where h stands for the increase in step size.

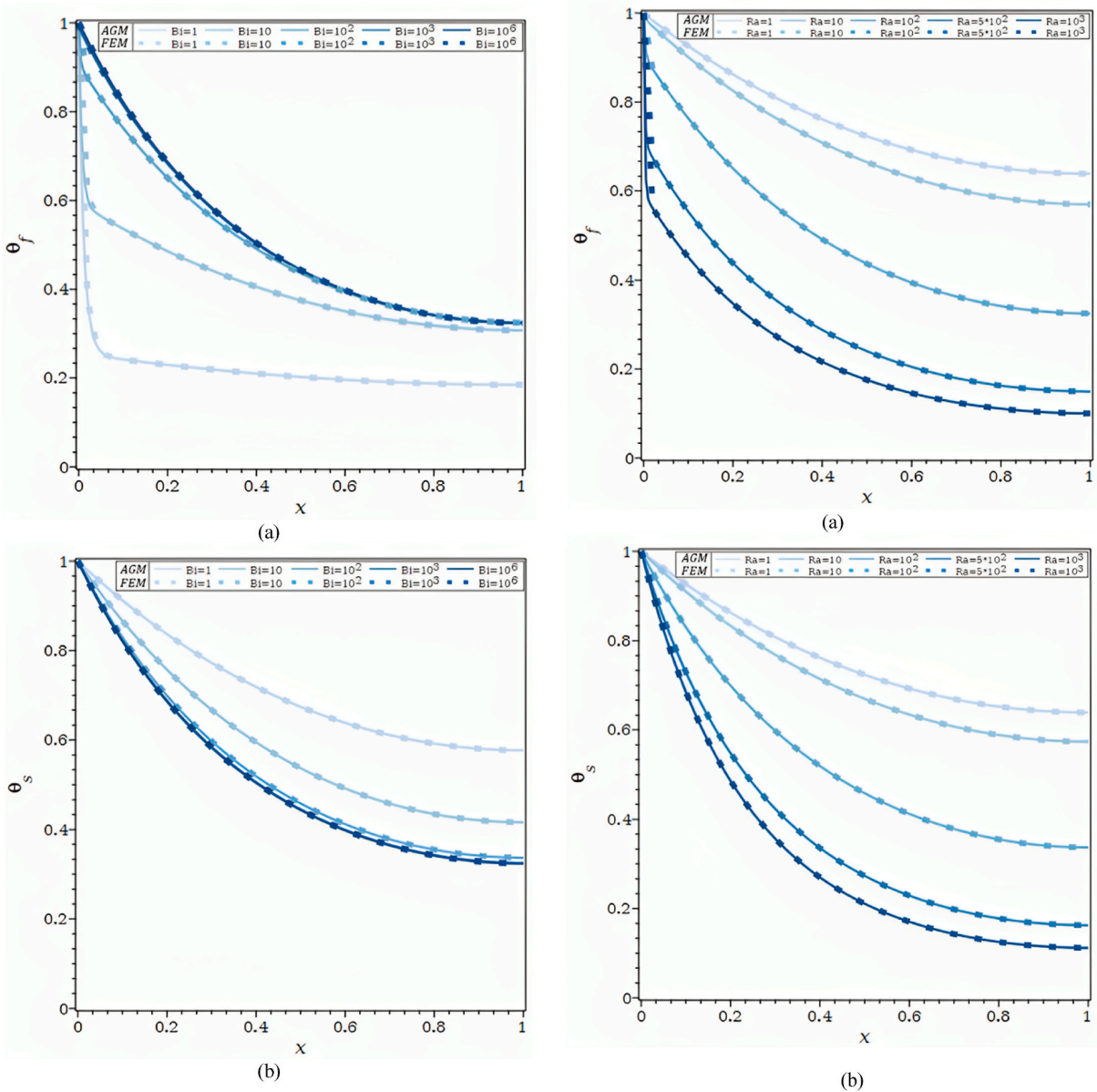

**Fig. 2.** a. Temperature profiles for different Bi values and Ra = 100, $\tau = 0.1$, $R_d = 0.3$, $\lambda = 0.01$, $R_1 = 0.1$, $\varphi = 0.92$, $\kappa = 10,000$: (a) $\theta_f$, (b) $\theta_s$.
Fig. 2b. Temperature profiles for different Ra values and Bi = 100, $\tau = 0.1$, $R_d = 0.3$, $\lambda = 0.01$, $R_1 = 0.1$, $\varphi = 0.92$, $\kappa = 10,000$: (a) $\theta_f$, (b) $\theta_s$.
Fig. 2c. $\theta_f$ and $\theta_s$ for different Ra values and $\tau = 0.1$, $R_d = 0.3$, $\lambda = 0.00$, $R_1 = 0.1$, $\varphi = 0.92$, $\kappa = 10,000$: (a) Bi = 1, (b) Bi = 100, (c) Bi = 1000.
Fig. 2d. $\theta_f$ and $\theta_s$ for different $\lambda$ values and Ra = 100, $\tau = 0.1$, $R_d = 0.3$, $R_1 = 0.1$, $\varphi = 0.92$, $\kappa = 10,000$: (a) Bi = 1, (b) Bi = 10, (c) Bi = 1000.

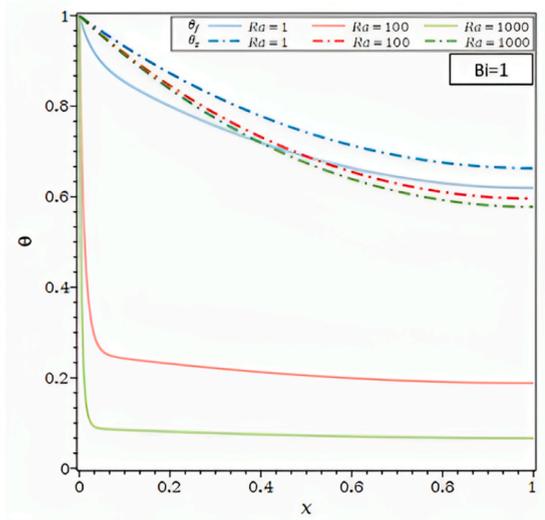

(a)

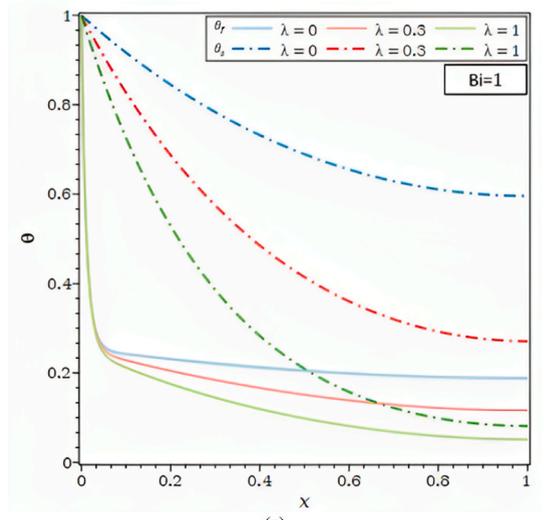

(a)

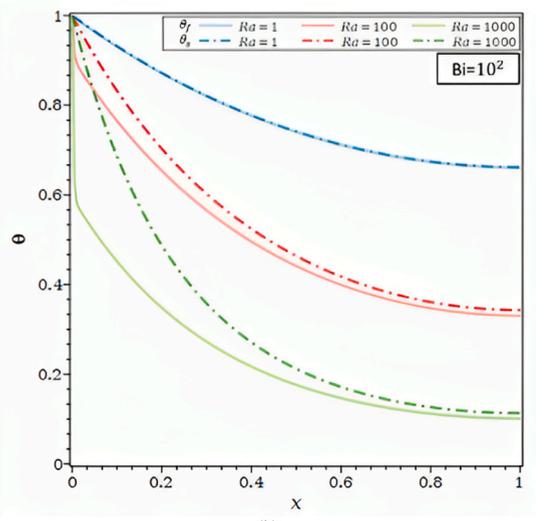

(b)

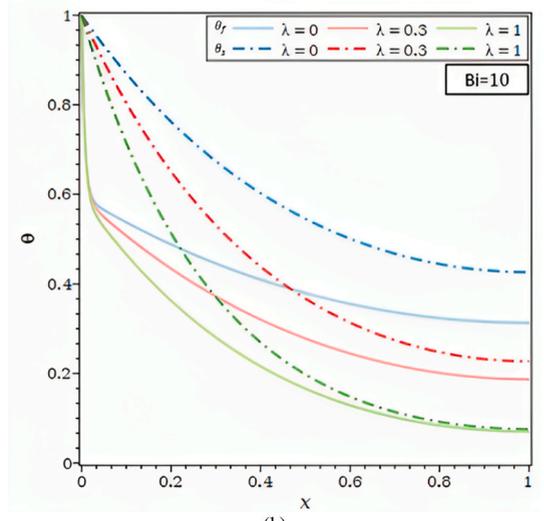

(b)

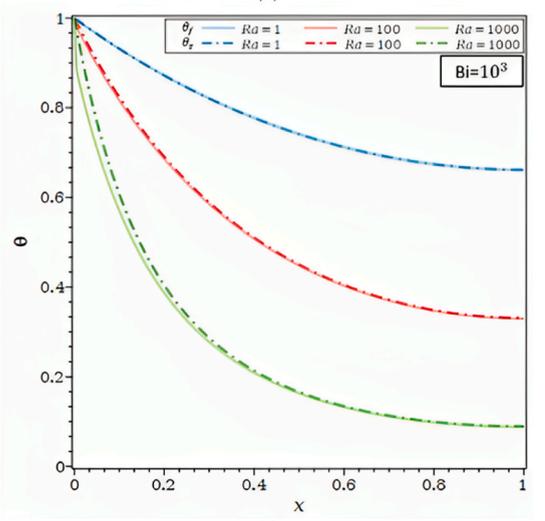

(c)

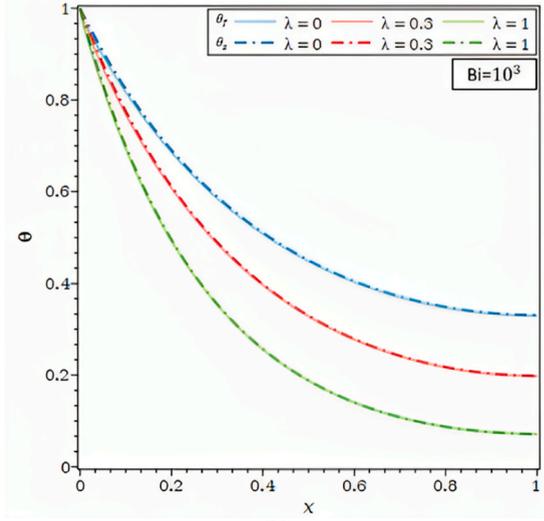

(c)

**Fig. 2.** (*continued*).

The function's second-order derivative at $x$ can be obtained as (Equation (36))

$$\Delta^2 \theta_i = \frac{d}{dx}\left(\frac{d\theta(x)}{dx}\right)\bigg|_{x=x_i} = \frac{\Delta\theta_{i+1} - \Delta\theta_i}{h} = \frac{\theta_{i+2} - 2\theta_{i+1} + \theta_i}{h^2}. \tag{36}$$

Higher-order derivatives can be derived using the same method.

### 3.3.2. Finite element method

Differential equations are discretized numerically using the finite element method to create algebraic equations. Compared to the finite difference method, FDM often considers the node spacing so that the entire domain is divided into squares and rectangles. The FEM gets around this problem by evenly spacing the nodes so the domain can be divided into shapes. The finite element method divides the global domain into a finite number of nonoverlapping subdomains.

### 3.4. Validation

Using the designated boundary conditions mentioned in Eqs. (14)–(16), the AGM used to assess the solution to the ODE mathematical problem is validated. The validation is carried out by comparing the results from Buonomo et al. [23] with the solutions obtained using Akbari Ganji, finite difference, and finite element methods.

The dimensionless fluid and solid phase temperature distributions obtained using various techniques are displayed in Fig. 1b and Table 1 for Bi = 1, Ra = 50, $\tau = 0.1$, $R_d = 0.1$, $\lambda = 0.01$, $R_1 = 0.3$, $\varphi = 0.92$, and $\kappa = 1000$. Due to the solid's greater thermal conductivity, Compared to solid temperature profiles, the convergence of fluid temperature profiles happens more slowly, which is why the fluid temperature profile requires a 23-term exponential (equation (29)) instead of the solid temperature profile's 7-term polynomial (equation (30)). It's intriguing to note that a high-temperature gradient occurs near the fin base because the fluid phase has less thermal conductivity than the solid phase.

In Fig. 1c and Table 2, the $\theta_f$ and $\theta_s$ for the values of Bi = 100, Ra = 50, $\tau = 0.1$, $R_d = 0.1$, $\lambda = 0.01$, $R_1 = 0.3$, $\varphi = 0.92$, $\kappa = 1000$ are shown. It should be observed that whereas Bi = 1 and 100 denote little and significant convective heat transfer, respectively, between the solid phase and fluid phase inside the porous fin, $\kappa = 1000$ denotes a very large differential in thermal conductivities.

For Bi = 1.0 and 100, shown in Fig. 1d, the Nusselt number as a function of Ra employing various techniques is presented in Table 3. The fluid enters the porous fin at $T_\infty$, and the $Nu$ number is the total amount of heat transferred via the porous fin from the surface at $T_b$ to the fluid at $T_\infty$.

## 4. Result and discussion

By assuming local thermal non-equilibrium between the fluid and solid phases, the findings of the Akbari-method Ganji's (AGM) utilized to solve the energy equations 14 and 15 for finite-length fins with adiabatic tips are illustrated and described below. In addition to an overall $Nu$ number, the findings are shown as dimensionless temperature profiles for the solid and fluid phases along porous fins.

Fig. 2a, which shows the temperature of the solid and fluid phases along the dimensionless axis x for different Bi, shows that convective heat transfer occurs at low to high relative to conduction rates. The values of $\kappa = 10^4$ and very high thermal conductivity differences result in a large temperature difference between the solid and fluid phases when air serves as the fluid and an aluminum fin is porous, there is relatively little external convective heat transfer, and Bi = 0. As the temperature profiles of the solid and fluid decrease down the axis x, as predicted, the conductive effects decrease along the axis of the porous fin. The conductive effect predominates when Bi = 1.0, and as a result, for high $\kappa = 10^4$, a high gradient temperature for the fluid phase is recorded at the base of the fin. The $\theta_f$ and $\theta_s$ are almost identical for $Bi \geq 10^3$. Therefore, the solid matrix and the fluid (LTE) have reached a local thermal equilibrium. The solid directly transfers heat, resulting in a global decrease in thermal resistance.

Fig. 2b illustrates the $\theta_f$ and $\theta_s$, along the permeable fin's axis for various values of the Rayleigh number. Temperature profiles of the fluid and solid matrix trend toward the ambient temperature more quickly as Ra rises. The porous fin makes it possible for more heat to be evacuated from the base where it is positioned. By noting that for certain temperature differences between the fin base, $T_b$, and ambient, $T_\infty$, the Rayleigh number, Ra, depends on the permeability of the porous material. Ra increases by increasing heat transfer by buoyancy as a result. Nearly no differences were seen between AGM and FEM for both fluid and solid profiles.

For varied Bi, Fig. 2c displays the Ra dependence of the fluid and solid temperature profiles. As the Rayleigh number, Ra, climbs, the temperature disparities between the fluid and solid in Fig. 2c in part (a) for Bi = 1 grow. As the Biot number rises, they get smaller, as seen in Fig. 2c parts (b) and (c). This issue demonstrates that significantly bigger Biot numbers are required for higher Rayleigh numbers to achieve local thermal equilibrium in both the solid and the fluid matrix.

In Fig. 2d, for Ra = $10^2$, Bi = 1, 10 and $10^3$, the impact of the heat transfer parameter, $\lambda$, on the $\theta$ is shown. The $\theta_f$ and $\theta_s$, in Fig. 2d, parts (a) and (b) drop when the external convective heat transfer parameter $\lambda$ rises. The drop in $\theta_s$ along the fin and, so, the fall in $\theta_f$, along the porous fin are caused by a rise in convective heat transfer to the fluid environment outside. Furthermore, in Fig. 2d part (c), the fluid and solid matrix both meet the local thermal equilibrium condition for Ra = $10^2$ and Bi = $10^3$.

Fig. 3a illustrates the impact of the internal radiation parameter, $R_d$, on the temperature profile within the fin for various Ra values. A rise in the internal radiative parameter increases the porous fin's fluid and solid temperatures for all Bi values. The internal radiation parameter has increased Since $R_d$, causes the porous fin's radiative-conductive resistance to be lower, the temperature along the fin is likely to be more consistent and approaching base temperature. Additionally, as the internal radiative parameter changes, the temperature differences between the fluid and solid phases are more pronounced for Bi = 10 than for Bi = 100, whereas for Bi = 10000, the $\theta_f$ and $\theta_s$, are nearly identical irrespective of $R_d$'s internal radiative parameter values. The LTE assumption is valid for all $R_d$ values are taken into account in the current study and a high Biot number value of $10^4$.

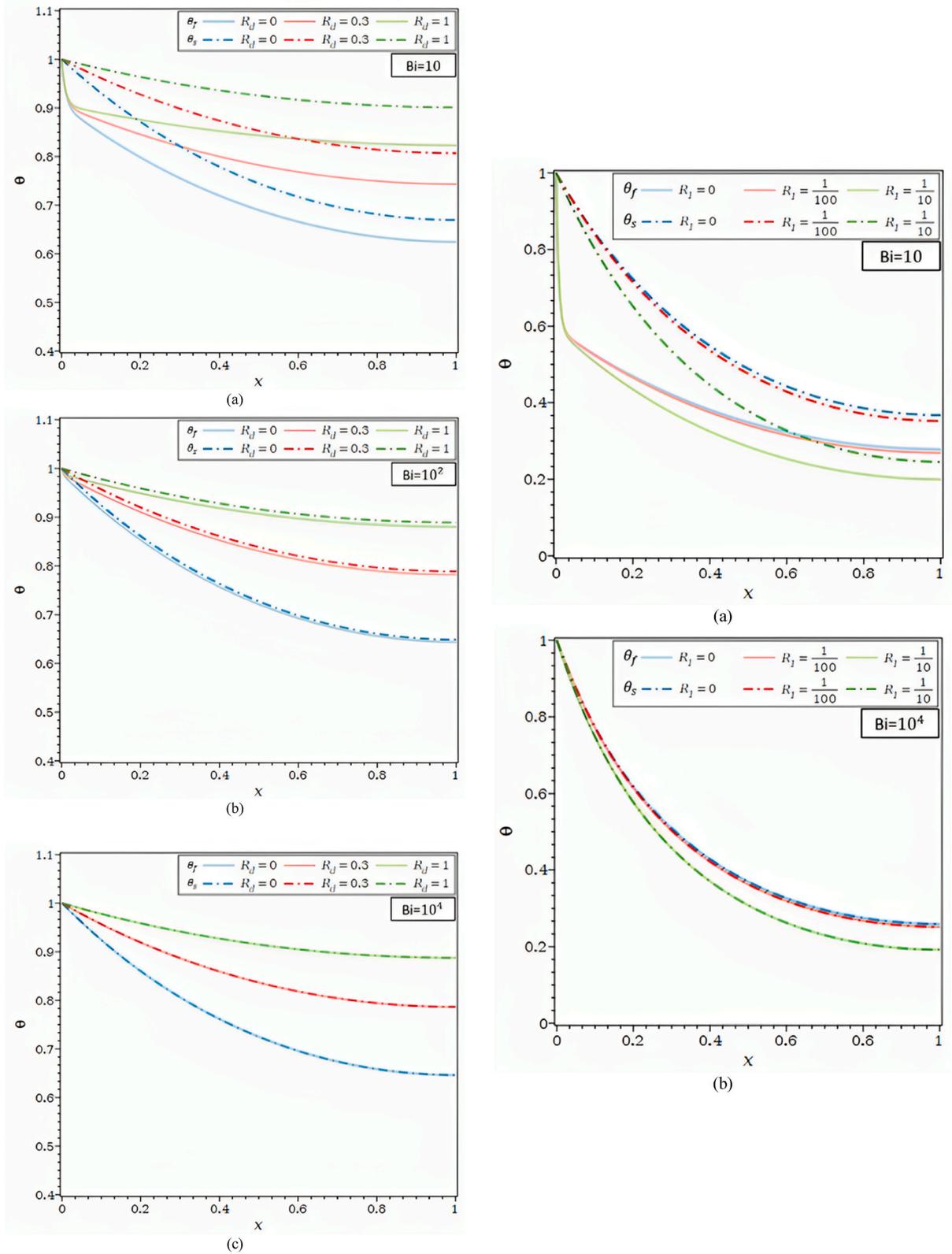

**Fig. 3.** a. $\theta_f$ and $\theta_s$ for different $R_d$ values and Ra = 10, $\tau = 0.1$, $\lambda = 0.01$, $R_1 = 0.0$, $\varphi = 0.92$, $\kappa = 10{,}000$: (a) Bi = 10, (b) Bi = 100, (c) Bi = 10,000.
Fig. 3b. $\theta_f$ and $\theta_s$ for different $R_1$ values and Ra = 100, $\tau = 0.1$, $R_d = 0.0$, $\lambda = 0.00$, $\varphi = 0.92$, $\kappa = 10{,}000$: (a) Bi = 10, (b) Bi = 10,000.

Fig. 3c. $\theta_f$ and $\theta_s$ for different $\kappa$ values and Ra = 100, $\tau$ = 0.1, $R_d$ = 0.0, $\lambda$ = 0.00, $R_1$ = 0.0, $\varphi$ = 0.92: (a) Bi = 10, (b) Bi = 10,000.

Fig. 3d. $\theta_f$ and $\theta_s$ for different $\kappa$ values and Bi = 100, $\tau$ = 0.1, $R_d$ = 0.0, $\lambda$ = 0.00, $R_1$ = 0.0, $\varphi$ = 0.92: (a) Ra = 1, (b) Ra = 100.

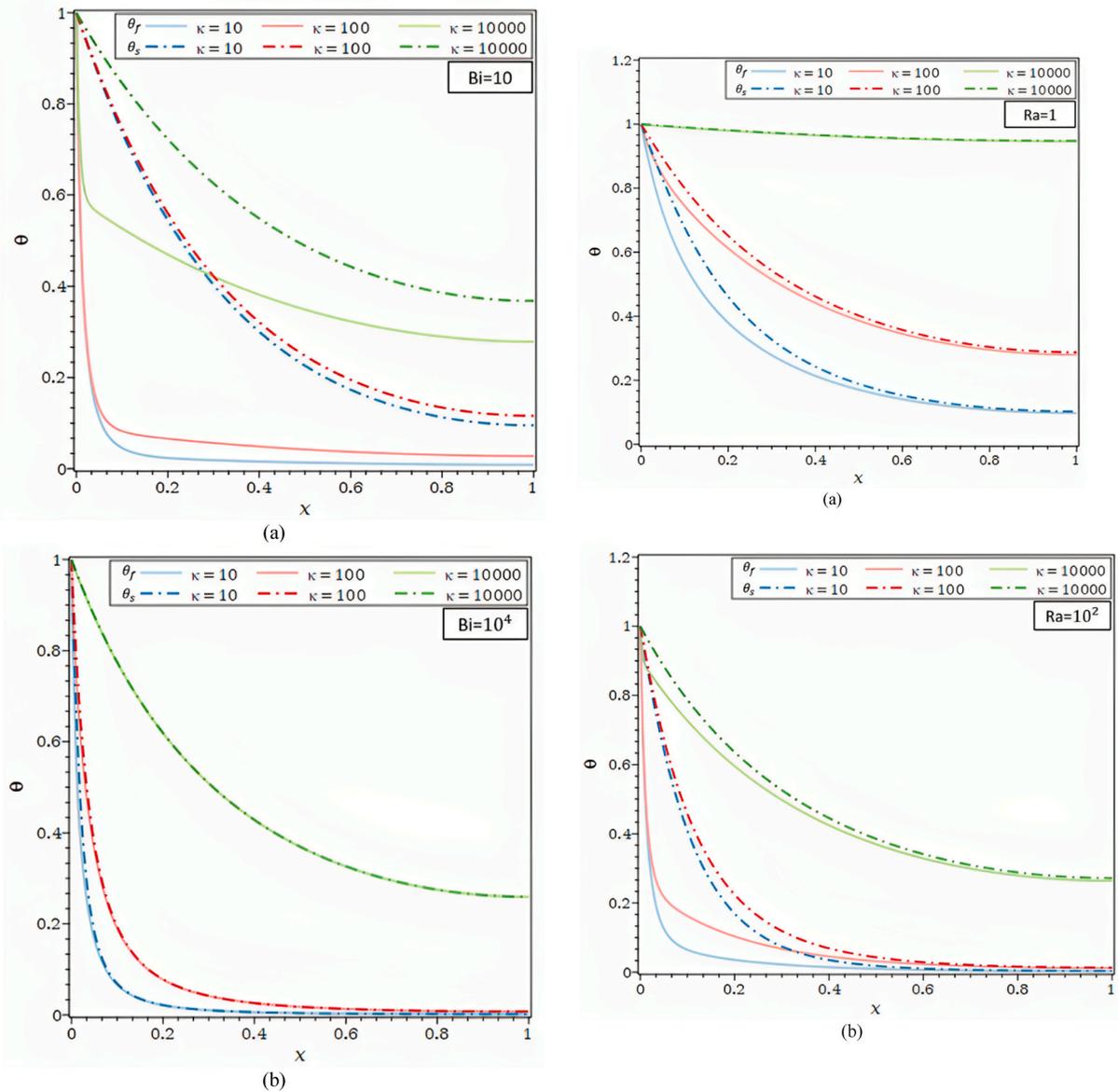

**Fig. 3.** (*continued*).

Fig. 3b depicts the impact of the external radiation parameter, $R_1$, on temperature distribution for Ra = $10^2$ and Bi = 10 and $10^4$. Fig. 3b shows that the temperature decreases as the external radiative parameter R1 rises. In actuality, as $R_1$ rises, the solid matrix's radiative heat transfer to the environment also rises; as a result, the $\theta_s$, along the fin falls along with the fluid temperature. The solid matrix and the fluid are likewise seen to be in a local thermal equilibrium condition for Ra = $10^2$ and Bi = $10^4$. The temperatures also fall by R1 and grow by the same proportion as the radiative parameter.

Fig. 3c reports the fluid temperature profiles and solid temperature profiles for various thermal conductivity ratios, including $\kappa$ = 10 (air/carbon foam), $\kappa = 10^2$ (water/SiC foam), and $\kappa = 10^4$ (air/Aluminum foam). All situations under consideration show that the $\theta_s$ and $\theta_f$, profiles increase. Increasing the thermal conductivity of the porous matrix increases the thermal conductivity ratio, which causes the solid's thermal conductivity resistance to drop. The increase in $\theta_s$ and, subsequently, that of the $\theta_f$, are determined by the decrease in conductive thermal resistance. As shown in Fig. 3c, temperature disparities between solid and fluid also Biot number rise and diminish as the thermal conductivity ratio $\kappa$.

As seen in Fig. 3d, the $\theta$ are shown for {Ra = 1, 100}.The differences between $\theta_f$ and $\theta_s$ for the value of the allocated Biot number $10^2$, and distinct thermal conductivity ratios grow with a rise in the Rayleigh number. The discrepancies between the $\theta_f$ and $\theta_s$ decrease noticeably for lower Ra values equivalent to one in Fig. 3d and vanish when the thermal conductivity ratios are pretty high, $\kappa = 10^4$. Lower k values lead to a decline in the diffusive effect compared to the buoyancy effect, which raises the temperature gradient along the fin. For the identical value of k, a more considerable Ra value of $10^4$ results in a higher temperature gradient inside the fin and a buoyancy effect more significant than the one in the preceding case. According to Fig. 3d, the temperature changes between the $\theta_f$ and $\theta_s$, tend to zero for thermal conductivity ratio values, Rayleigh numbers, and X > 0.8.

The influence of the geometric parameter ($\tau = t/L$) on the temperature profiles of the solid and fluid is depicted in Fig. 4a, respectively, for Bi = 10 and Bi = $10^4$. $\tau = t/L$ might be decreased by either decreasing the thickness, t, for the provided fin length, L, or increasing the fin length, L. The fin tip's temperature grows along with the geometric parameter $\tau$ in Fig. 4a. The temperature rises noticeably the shorter the fin, and the higher temperature is near the base for a given fin thickness. Furthermore, when the temperature differential between the solid and fluid grows in Fig. 4a, for Bi = 10, the fin aspect ratio decreases. For a given fin length, an increase in the fin aspect ratio causes the thickness of the fin to grow while its thermal resistance decreases. As a result, the temperature gradient in both the solid and fluid phases is reduced as it moves toward the base of the fin.

Fig. 4b shows how the porosity parameter ($\varphi$) affects the $\theta_f$ and $\theta_s$ for Bi = 10 and Bi = $10^4$. The temperature at the fin tip is seen to decrease dramatically as the porosity parameter increases. Additionally, the difference in temperature between a fluid and a solid grows as the porosity parameter rises in Fig. 4b for Bi = 10. Furthermore, it can be seen that when $\varphi = 0.01$, the temperature differential between the $\theta_f$ and $\theta_s$ is insignificant, indicating that the LTNE assumption can be disregarded for nearly non-porous fins.

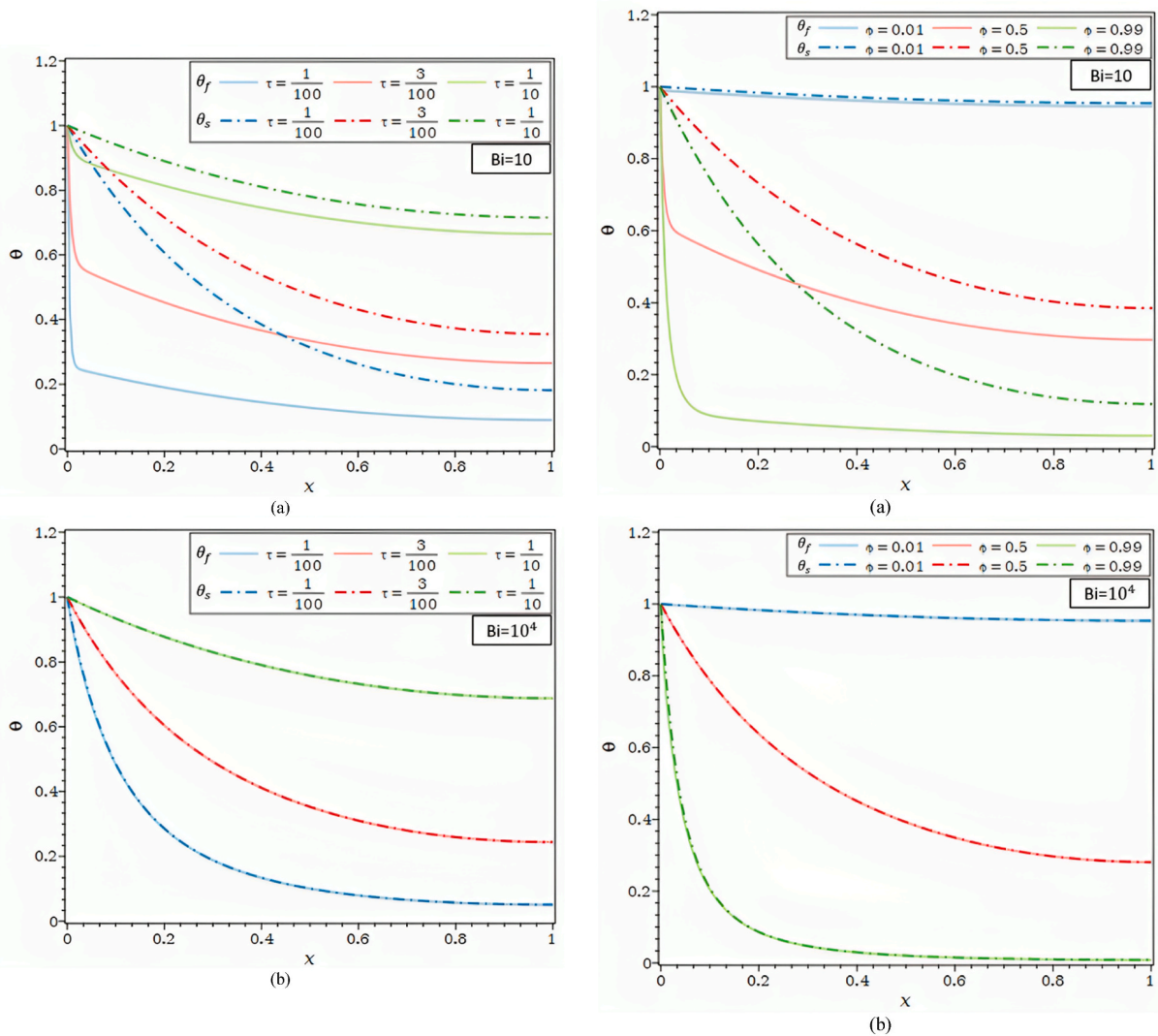

**Fig. 4.** a. $\theta_f$ and $\theta_s$ for different $\tau$ values and Ra = 10, $R_d$ = 0.0, $\lambda$ = 0.00, $R_1$ = 0.0, $\varphi$ = 0.92, $\kappa$ = 10,000: (a) Bi = 10, (b) Bi = 10,000.
Fig. 4b. $\theta_f$ and $\theta_s$ for different $\varphi$ values and Ra = 100, $\tau$ = 0.1, $R_d$ = 0.0, $\lambda$ = 0.00, $R_1$ = 0.0, $\kappa$ = 1000: (a) Bi = 10, (b) Bi = 10,000.
Fig. 4c. The overall $Nu$ number as a function of Ra for $\lambda$ = 0.00, $R_1$ = 0.0, $\varphi$ = 0.92: (a) different values of Bi, (b) different values of $\kappa$, (c) different values of $R_d$, (d) different values of $\tau$.

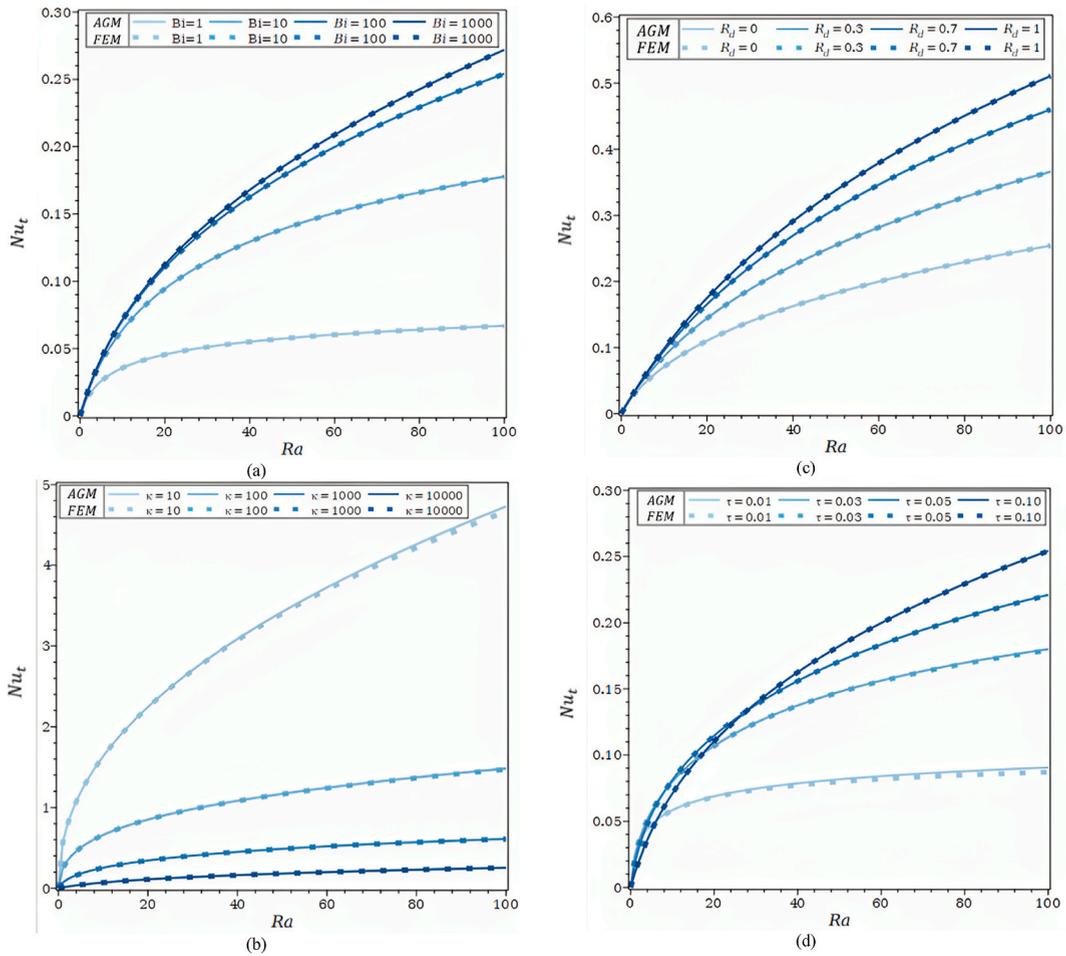

**Fig. 4.** (*continued*).

For ranges with dimensionless parameters of $1 \leq \text{Bi} \leq 10^3$ in Fig. 4c part (a), $10 \leq \kappa \leq 10^4$ in Fig. 4c part (b), $0 \leq R_d \leq 1$ in Fig. 4c part (c), and $0.01 \leq \tau \leq 0.1$ in Fig. 4c part (d), the total $Nu$ number profiles as a function of the Rayleigh number are presented.

The values of the $Nu$ number rise as the Biot number rises, as seen in Fig. 4c. The $Nu_t$ profiles converge toward an asymptotic profile compatible with the local thermal equilibrium condition that exists for Bi = 10,000 as the Biot number rises. When the Biot number and convective heat transfer coefficient at the solid matrix-fluid interface expand, the rate at which heat is evacuated from the base of the fin also climbs. The effect of the fluid-solid thermal conductivity ratio is seen in Fig. 4a, $\kappa$, on the total Nusselt number when $\varphi = 0.92$, Bi = 100, $\tau = 0.1$, and $R_d = R_1 = \lambda = 0$. For a given Ra value, it has been found that the total Nusselt number drops as the thermal conductivity ratio rises. When with metal fins, the temperature of the fin tends to the base temperature as the thermal resistance of the solid matrix falls and the ratio of thermal conductivities between fluid and solid rises (Al, Cu). Because $R_d = R_1 = \lambda = 0$, the heat transfer rate from the porous matrix drops.

Fig. 4c part (d) shows the impact of the porous fin's aspect ratio on the total $Nu$ number, $Nu_t$, for the values of Bi = 100, $\kappa = 10000$, $\varphi = 0.92$ and $R_d = R_1 = \lambda = 0$. For $\tau = 0.1$, the $Nu_t$ is lower than for $\tau = 0.01$ up to around Ra = 6. Concerning the value for $\tau = 0.1$, the trend reversal for $\tau = 0.03$ and 0.05 begins when Ra is approximately (18–27), respectively. Ra values above 27 cause the temperature differential between the base and ambient to grow, and shorter fins exhibit higher temperatures. Therefore, the rate of heat transmission rises as the value rises. The various trends can be explained if the porous fin's heat transmission rate is considered. The difference in temperature between the base of the fins and the ambient has to do with the external heat transfer surface. The most important idea is the heat transmission area, indicated by low Ra or assigned thickness, which suggests little variance in base and ambient temperatures. The heat transfer area, A, and fin length, L, are high for a low dimensionless thickness. It is important to notice that the total $Nu$ number tends to an asymptotic value of 0.08 for $\tau = 0.01$. As a matter of fact, for $\tau = 0.01$, the fin approaches the infinitely long fin ($\tau \to 0$). It has no heat transmission and a huge region with a temperature similar to the ambient. The $Nu$ number does not approach an asymptotic value for the chosen Ra values and aspect ratio values below 0.03. There was a minimal distinction between AGM and FEM in the overall $Nu$ number profiles as a function of the Rayleigh number. One of the things that can be researched in the future is the investigation of the effect of the magnetic field on the flow and heat fields. According to the use of hybrid Nanos in recent years, the effect of using hybrid nanofluids to improve heat transfer is one of the new topics that can be researched.

## 5. Conclusion

Under local thermal non-equilibrium, a straightforward thermal model of a porous fin with an adiabatic tip and a limited length was proposed. The first study of two energy equations for permeable fins was performed using the Akbari Ganji-approach. Examining the temperature profiles of the solid and fluid phases along the porous fin and the overall $Nu$ number in terms of thermal factors and geometrical emphasized thermal phenomena. As expected, the lowest Bi value showed the most significant disparities between the temperature profiles of solids and fluids. Near the fin's base, the most significant variances were seen. The impacts of $R_a$ for the given Bi value were considerable, and for low Bi = 1, the percentage variances between the $\theta_s$ and $\theta_f$ at the tip of the fin, referred to as the base temperature, were 52% and 3% for Ra = $10^3$ and 1, respectively. For Ra = 1000 and 1, the values for Bi = 100 were 1% and almost 0. Lower discrepancies between the $\theta_s$ and $\theta_f$, were found for the prescribed Bi value by the rise in external convection.

For the provided values of Bi and Ra, the increase in the thermal conductivity ratio led to a decrease in the solid phase's thermal resistance, an increase in the temperature of the solid phase, and a rise in the gaps between the solid and fluid phases' temperature profiles. Ra and Bi increments result in a rise in the total Nusselt number or heat removed from the porous fin. For Bi = $10^4$, $Nu_t$, gave an asymptotic profile. In comparison to the LTNE assumption, the LTE assumption results in an overestimation of $Nu_t$.

### Data availability statement

Data will be made available on request.

### Additional information

No additional information is available for this paper.

### CRediT authorship contribution statement

**Payam Jalili:** Conceptualization. **Salar Ghadiri Alamdari:** Writing – original draft, Software. **Bahram Jalili:** Validation, Supervision. **Amirali Shateri:** Writing – review & editing. **D. D. Ganji:** Supervision.

### Declaration of competing interest

The authors declare that they have no known competing financial interests or personal relationships that could have appeared to influence the work reported in this paper.

### Nomenclature

| | |
|---|---|
| $a_{sf}$ | interfacial area per unit of volume of porous fin, $m^{-1}$ |
| $A_b = Wt$ | the cross-section area of the fin, $m^{-2}$ |
| $A_{sf}$ | the interfacial area between fluid and solid phases, $m^2$ |
| $Bi$ | Biot number |
| $c_{pf}$ | specific heat of fluid, $Jkg^{-1}K^{-1}$ |
| $g$ | acceleration of gravity, $ms^{-2}$ |
| $h_L$ | external convective heat transfer coefficient, $Wm^{-2}K^{-1}$ |
| $h_{sf}$ | convective heat transfer coefficient between solid and fluid phases in porous fin, $Wm^{-2}K^{-1}$ |
| $k$ | thermal conductivity, $Wm^{-1}K^{-1}$ |
| $K$ | permeability of porous fin, $m^2$ |
| $L$ | length of the porous fin, m |
| $\dot{m}$ | mass flow rate, $kgs^{-1}$ |
| $Nu$ | Nusselt number |
| $P$ | cross-section perimeter, m |
| $q$ | heat transfer flux, W $m^2$ |
| $q_x$ | conductive heat transfer flux component, along x, W $m^2$ |
| $\dot{Q}$ | heat transfer rate |
| $R_1$ | surface-ambient radiation parameter |
| $Ra$ | Rayleigh number for porous medium |

| $R_d$ | radiation–conduction parameter |
|---|---|
| $t$ | the thickness of porous fin, m |
| $T_\infty$ | ambient temperature, K |
| $T_b$ | the temperature of the porous fin base, K |
| $T$ | temperature, K |
| $x$ | axial coordinate, m |
| $X$ | dimensionless axial coordinate |
| $v_w$ | velocity in the porous fin, $ms^{-2}$ |
| $W$ | width of the porous fin, m |

*Greek symbols*

| $\beta$ | coefficient of volumetric expansion, $K^{-1}$ |
|---|---|
| $\beta_R$ | Rosseland mean extinction coefficient, m |
| $\nu$ | kinematic viscosity, $m^2 s^{-1}$ |
| $\varphi$ | porosity |
| $\varepsilon$ | emissivity coefficient |
| $\theta$ | dimensionless temperature |
| $\chi$ | dimensionless parameter |
| $\kappa$ | thermal conduction ratio |
| $\tau$ | dimensionless thickness |
| $\lambda$ | external convective heat transfer coefficient |
| $\sigma$ | Stefan-Boltzmann constant, $Wm^{-2}K^{-4}$ |
| $\rho$ | density, $kgs^{-3}$ |

*Subscripts*

| $b$ | base of fin |
|---|---|
| $f$ | fluid |
| $eq$ | effective |
| $l$ | effective |
| $L$ | system length scale, i.e., porous fin length |
| $s$ | system length scale, i.e., porous fin length |
| $s,f$ | solid-fluid interface |
| $t$ | total |